\documentclass [12pt]{article}
\begin{document}

\title{\textbf{Density expansion of the
energy \\ of N close-to-boson excitons}}
\author{O. Betbeder-Matibet and M.
Combescot \\GPS, Universit\'e Denis Diderot
and Universit\'e Pierre et Marie Curie,
CNRS,\\Tour 23, 2 place Jussieu, 75251
Paris Cedex 05, France}
\date{}
\maketitle

\begin{abstract}

Pauli exclusion between the carriers of $N$ excitons
induces novel many-body effects, quite different from
the ones generated by Coulomb interaction. Using our
commutation technique for interacting close-to-boson
particles, we here calculate the hamiltonian
expectation value in the $N$-ground-state-exciton state.
Coulomb interaction enters this quantity at first order
only by construction ; nevertheless, due to Pauli
exclusion, subtle many-body effects take place, which
give rise to terms in
$(Na_x^3/\mathcal{V})^n$ with
$n\geq2$. An \emph{exact} procedure to get
these density dependent terms is given. 
\end{abstract}

\vspace{2cm}


\newpage

It is known that excitons being made of two
fermions are not exact bosons. Their
underlying fermionic character is however a
major difficulty which has been overcomed
very approximately up to quite recently. An
approach $^{(1,2)}$ which has looked reasonable for
years is to assume that excitons are exact
bosons provided that their close-to-boson
character is included in an appropriate
exciton-exciton interaction which is
basically a Coulomb interaction dressed by
exchange processes.

Very recently, we have developed a formalism
which allows to
include the close-to-boson character of the
excitons \emph{exactly}, through an extremely
simple and physically quite transparent
algebra $^{(3)}$. Using this "commutation technique",
we have already shown $^{(4)}$, by calculating
the correlations of \emph{two} excitons, that
Pauli exclusion between the two electrons and
the two holes of these two excitons, enters
their Coulomb terms in such a subtle way that
a na\"{\i} ve bosonic hamiltonian for excitons $^{(1,2,5)}$
cannot produce these terms correctly beyond
first order in Coulomb interaction, reducing
considerably the impact of such an effective
hamiltonian.

A way to physically understand the problem is
to realize that excitons feel each other
through both Coulomb interaction and Pauli
exclusion between their components.
Consequently, besides the usual many-body
effects resulting from Coulomb interaction,
the excitons do have very unusual ones
coming from Pauli exclusion. These two kinds
of many-body effects being basically
independent, they enter the various
quantities of physical interest quite
differently. So that there is no
reason, for the exchange processes resulting
from Pauli exclusion, to dress the Coulomb
interaction in an unique way. In other words,
there is no reason to describe the physics of
interacting excitons by \emph{one} dressed
exciton-exciton interaction only, as assumed
in the effective bosonic hamiltonian.

In the present work, we calculate the
semiconductor hamiltonian expectation value in
the
$N$-ground-state-exciton state. We here forget about the spin
degrees of freedom $^{(6)}$ of the excitons, for the sake of
simplicity. In the absence of Coulomb interaction and Pauli
exclusion, this hamiltonian expectation value should be
equal to $NE_0$ with
$E_0$ being the ground state energy of one exciton.
Due to Coulomb interaction, which by construction enters
this quantity at first order only, we should find
\emph{one} additional term resulting from the
average Coulomb energy in this $N$-exciton
state, and which should be $Na_x^3/\mathcal{V}$ smaller
than the main energy $NE_0$, $a_x$ being the
exciton Bohr radius and $\mathcal{V}$ the sample volume.
The exact calculation shows that the
hamiltonian expectation value also contains
terms in $(Na_x^3/\mathcal{V})^n$ with $n³2$, which
originate from many-body effects between the
$N$ excitons induced by Pauli exclusion.
Using our commutation technique, we show how
we can derive this $N$-exciton average
energy at any order in $\eta=Na_x^3/\mathcal{V}$ in a
systematic way.

The paper is organized as follows. In section 1, we recall
the basic relations of our commutation technique and use them
to calculate the expectation value of the
hamiltonian in the $N$-ground-state-exciton state, $<H>_N$.
In section 2, we derive the recursion relation satisfied by
the matrix elements which appear in the expression of
$<H>_N$, and we iterate this relation to obtain the density
leading terms of these matrix elements. In section 3, we
calculate the three first terms of the
$\eta=Na_x^3/\mathcal{V}$ expansion of
$<H>_N$ explicitly. In the conclusion, we discuss the
obtained result and relate it to the new criterion for
bosonic behavior of excitons we recently proposed $^{(7)}$. 

\section{Expectation value of the hamiltonian using our
commutation technique}
 The semiconductor hamiltonian expectation
value in the $N$-ground-state-exciton state
reads
\begin{equation}
<H>_N=\frac{<v|(B_0)^N H
(B_0^\dag)^N|v>}{<v|(B_0)^N(B_0^\dag)^N|v>}\ 
,
\end{equation}
where $|v>$ is the electron-hole pair vacuum
state and $B_0^\dag$ the (exact) creation operator of
one ground state exciton, i.\ e.\ exciton with a center of
mass momentum $\mathbf{Q}_0=\mathbf{0}$ and a relative
motion in its ground state $\varphi_{\nu_0}$. The algebraic
calculation of
$<H>_N$ is quite easy through our commutation technique. 

Two important relations of this commutation
technique are $^{(3)}$
\begin{equation}
\left[H,B_i^\dag\right]=E_iB_i^\dag
+V_i^\dag\ ,
\end{equation}
\begin{equation}
\left[V_i^\dag,B_j^\dag\right]=\sum_{mn}\xi_{mnij}^\mathrm{dir}B_m^\dag
B_n^\dag\ ,
\end{equation}
where $E_i$ is the $i$ exciton energy and $B_i^\dag$ the
(exact) creation operator of an $i$ exciton, $i$ standing for
$(\nu_i,\mathbf{Q}_i)$. The operator
$V_i^\dag$ comes from the Coulomb interactions between the $i$ exciton
and the rest of the system, while the 
parameter $\xi_{mnij}^{\mathrm{dir}}$
corresponds to the direct Coulomb scattering
of the $(i,j)$ excitons into the $(m,n)$
states. In $\mathbf{r}$ space, $\xi_{mnij}^{\mathrm{dir}}$ 
reads \nolinebreak $^{(3)}$ 
\begin{eqnarray}
\xi_{mnij}^{\mathrm{dir}}=\xi_{ijmn}^{\mathrm{dir}\ \ast}&=&\frac{1}{2}\int
d\mathbf{r}_{e_1}\,d\mathbf{r}_{h_1}\,d\mathbf{r}_{e_2}\,d\mathbf{r}_{h_2}\,\phi_m^\ast(e_1,h_1)\phi_n^\ast(e_2,h_2)\nonumber
\\ &\times& \left[V_{e_1e_2}+V_{h_1h_2}-V_{e_1h_2}-V_{e_2h_1}\right]
\phi_i(e_1,h_1)\phi_j(e_2,h_2)+(i\leftrightarrow
j)\ ,
\end{eqnarray}
where $\phi_i(e,h)$ is the total wave
function of the $i$ exciton, i.\ e.\ the
product of the relative motion wave function
$\varphi_{\nu_i}(\mathbf{r}_e-\mathbf{r}_h)$, and the center
of mass wave function
$e^{i\mathbf{Q}_i.(\alpha_e\mathbf{r}_e+\alpha_h\mathbf{r}_h)}
/\sqrt{\mathcal{V}}$, with
$\alpha_e=1-\alpha_h=m_e/(m_e+m_h)$. 

Using Eqs (2,3), it is straightforward to
prove by induction that
\begin{equation}
\left[H,(B_0^\dag)^N\right]=NE_0(B_0^\dag)^N+\frac{N(N-1)}{2}\sum_{mn}\xi_{mn00}^{\mathrm{dir}}B_m^\dag
B_n^\dag
(B_0^\dag)^{N-2}+N(B_0^\dag)^{N-1}V_0^\dag\ ,
\end{equation}
so that, as $H|v>=0=V_0^\dag |v>$, we get with $H$ acting on
the left
\begin{equation}
<H>_N=NE_0+\frac{N(N-1)}{2}\sum_{mn}\xi_{00mn}^{\mathrm{dir}}\frac{<v|(B_0)^{N-2}B_mB_n(B_0^\dag)^N
|v>}{<v|(B_0)^N(B_0^\dag)^N|v>}\ .
\end{equation}
As expected, Coulomb interaction enters
$<H>_N$ at first order only, as $<H>_N$ is
linear in the Coulomb scatterings
$\xi_{00mn}^{\mathrm{dir}}$. 

If the excitons were exact bosons, the ratios in
the sum would be 1 for $m=n=0$, and 0
otherwise, so that we would get 
$<H>_N=NE_0+N(N-1)\xi_{0000}^{\mathrm{dir}}/2$.
We can then note that
for $n=i=j$, Eq (4) reduces to
\begin{eqnarray}
\xi_{miii}^{\mathrm{dir}}&=&\frac{1}{\mathcal{V}^2}\int
d\mathbf{r}_{e_1}\,d\mathbf{r}_{h_1}\,d\mathbf{r}_{e_2}\,
d\mathbf{r}_{h_2}\,
e^{i(\mathbf{Q}_i-\mathbf{Q}_m).(\alpha_e\mathbf{r}_{e_1}+
\alpha_h\mathbf{r}_{h_1})} \varphi_{\nu_m}^\ast
(\mathbf{r}_{e_1}-\mathbf{r}_{h_1})\,\varphi_{\nu_i}
(\mathbf{r}_{e_1}-\mathbf{r}_{h_1}) \nonumber 
\\ &\times&
\left|\varphi_{\nu_i}(\mathbf{r}_{e_2}-\mathbf{r}_{h_2})
\right|^2
\left(\frac{e^2}{|\mathbf{r}_{e_1}-\mathbf{r}_{e_2}|}
+\frac{e^2}{|\mathbf{r}_{h_1}-\mathbf{r}_{h_2}|}
-\frac{e^2}{|\mathbf{r}_{e_1}-\mathbf{r}_{h_2}|}
-\frac{e^2}{|\mathbf{r}_{e_2}-\mathbf{r}_{h_1}|}\right)
\nonumber
\\&=&0\ ,
\end{eqnarray}
as can be seen by exchanging $\mathbf{r}_{e_2}$ and
$\mathbf{r}_{h_2}$.
Consequently, strangely enough, if we forgot about the
close-to-boson character of the excitons, $<H>_N$ would be
equal to $NE_0$ as if the excitons were not interacting at
all !

Of course, excitons are not exact bosons. Let us now show how we can calculate the
matrix elements appearing in the sum of Eq (6) for exact excitons. Here again, two
important relations of our commutation technique make their calculation quite easy,
namely $^{(3)}$
\begin{equation}
\left[B_i,B_j^\dag\right]=\delta_{ij}-D_{ij}\ ,
\end{equation}
\begin{equation}
\left[D_{mi},B_j^\dag\right]=2\sum_n \lambda_{mnij}B_n^\dag\ .
\end{equation}
$D_{ij}$ is the deviation-from-boson operator, while the parameter $\lambda_{mnij}$
describes the fact that there are two ways to couple two
electrons $(e_1,e_2)$ and two holes $(h_1,h_2)$ to make two
excitons, either $(e_1,h_1)(e_2,h_2)$ or
$(e_1,h_2)(e_2,h_1)$. In
$\mathbf{r}$ space $\lambda_{mnij}$ reads
$^{(3)}$
\begin{eqnarray}
\lambda_{mnij}=\lambda_{ijmn}^\ast=\frac{1}{2} \int
d\mathbf{r}_{e_1}\,d\mathbf{r}_{h_1}\,d\mathbf{r}_{e_2}\,d\mathbf{r}_{h_2}\,\phi_m^\ast(e_1,h_1)\phi_n^\ast(e_2,h_2)\phi_i(e_1,h_2)\phi_j(e_2,h_1)\nonumber
\\+(i\leftrightarrow j)\ .
\end{eqnarray}
Eqs (8-10) are the key equations which allow an easy
calculation of the matrix elements appearing in $<H>_N$.

\section{Calculation of the matrix elements appearing in
$<H>_N$}
 From Eqs (8,9), it is straightforward to prove by
induction that
\begin{equation}
\left[B_m,(B_0^\dag)^N\right]=N\delta_{m0}(B_0^\dag)^{N-1}-N(N-1)\sum_p\lambda_{mp00}B_p^\dag(B_0^\dag)^{N-2}-
N(B_0^\dag)^{N-1}D_{m0}\ .
\end{equation}

Before calculating the matrix elements of $<H>_N$, with two
excitons possibly outside the ground state, let us first
consider the simpler ones, with only one exciton outside the
ground state, namely
\begin{equation}
\mathcal{A}_m^{(N)}=<v\mid(B_0)^{N-1}B_m(B_0^\dag)^N\mid v>/N!\ .
\end{equation}
As $D_{ij}\mid v>=0$, we get from Eq (11) that
$\mathcal{A}_m^{(N)}$ verifies the recursion relation
\begin{equation}
\mathcal{A}_m^{(N)}=\delta_{m0}\mathcal{A}_0^{(N-1)}
-(N-1)\sum_p\lambda_{mp00}\left(\mathcal{A}_p^{(N-1)}
\right)^\ast\
.
\end{equation}
If we iterate this equation, and note that
$\mathcal{A}_0^{(N)}$ is nothing but the quantity $F_N$ 
defined in Ref.\ (7), 
\begin{equation}
\mathcal{A}_0^{(N)}=F_N=<v\mid(B_0)^N(B_0^\dag)^N\mid v>/N!\
,
\end{equation}
we find that $\mathcal{A}_m^{(N)}$ expands as
\begin{eqnarray}
\mathcal{A}_m^{(N)}=F_{N-1}\delta_{m0}-(N-1)F_{N-2}\lambda_{m000}+(N-1)(N-2)F_{N-3}\sum_p\lambda_{mp00}\lambda_{00p0}\nonumber\\
-(N-1)(N-2)(N-3)F_{N-4}\sum_{pq}\lambda_{mp00}\lambda_{00pq}\lambda_{q000}+\cdots\
.
\end{eqnarray}
By expliciting the $\lambda$'s, it is possible to show
that, for
$m=0$, Eq (15) is nothing but the recursion relation between
the $F_N$'s given in Ref.\ (7), namely
\begin{equation}
F_N=F_{N-1}-(N-1)\,\sigma_2\,F_{N-2}+(N-1)(N-2)\,\sigma_3\,F_{N-3}-\cdots\ ,
\end{equation}
with 
\begin{eqnarray}
\sigma_n&=&\sum_{\mathbf{k}}|\Phi_{\nu_0}(\mathbf{k})|^{2n}\nonumber
\\&=&
\frac{16(8n-5)!!}{(8n-2)!!}\ (64\pi a_x^3/\mathcal{V})^{n-1}\
, 
\end{eqnarray}
in 3D, $\Phi_{\nu_0}(\mathbf{k})$, given in the
appendix, being the Fourier
transform of $\varphi_{\nu_0}(\mathbf{r})$.

Let us now turn to the matrix elements with two excitons
possibly outside the ground state, namely
\begin{equation}
\mathcal{B}_{mn}^{(N)}=<v\mid (B_0)^{N-2}B_mB_n(B_0^\dag)^N\mid v>/N!\ .
\end{equation}
From Eq (9), we get
\begin{equation}
\left[D_{mi},(B_0^\dag)^N\right]=2N\sum_q\lambda_{mqi0}B_q^\dag(B_0^\dag)^{N-1}\ ,
\end{equation}
easy to prove by induction. Using Eqs (11,18,19), we
find that $\mathcal{B}_{mn}^{(N)}$ verifies the recursion
relation,
\begin{eqnarray}
\mathcal{B}_{mn}^{(N)}&=&F_{N-2}(\delta_{m0}\delta_{n0}
-\lambda_{mn00})\nonumber\\&-&(N-2)\sum_q
\left(\mathcal{A}_q^{(N-2)}\right)^\ast\left\{
\left(\delta_{m0}\lambda_{nq00}-\sum_p\lambda_{mp00}
\lambda_{nqp0}\right)+(m\leftrightarrow n)\right\}\nonumber
\\&+&(N-2)(N-3)\sum_{pq}\left(\mathcal{B}_{pq}^{(N-2)}\right)
^\ast\lambda_{mp00}\lambda_{nq00}\
.
\end{eqnarray}
We can then use Eq (15) for
$\mathcal{A}_q^{(N)}$ and iterate Eq (20). The three first
terms of the expansion of $\mathcal{B}_{mn}^{(N)}$ are
obtained by replacing $\mathcal{A}_q^{(N-2)}$ by the two
first terms of Eq (15) and $\mathcal{B}_{pq}^{(N-2)}$ by the
first term of Eq (20). This leads to
\begin{eqnarray}
\mathcal{B}_{mn}^{(N)}=F_{N-2}(\delta_{m0}\delta_{n0}
-\lambda_{mn00})\hspace{9.5cm}\nonumber\\
-(N-2)F_{N-3}\left\{\
\left(\delta_{m0}\lambda_{n000}-\sum_p\lambda_{mp00}
\lambda_{n0p0}\right)+(m\leftrightarrow
n)\ \right\}\hspace{3.4cm}\nonumber\\ 
+(N-2)(N-3)F_{N-4}\left\{\
\left[
\left(\delta_{m0}\sum_p\lambda_{np00}\lambda_{00p0}
-\sum_{pq}\lambda_{mp00}\lambda_{nqp0}\lambda_{00q0}\right)
+(m\leftrightarrow n)\right]\right.\nonumber
\\
\left.+\lambda_{m000}\lambda_{n000}-\sum_{pq}
\lambda_{mp00}\lambda_{nq00}\lambda_{00pq}\
\right\} +\cdots
\end{eqnarray}
The first term of $\mathcal{B}_{mn}^{(N)}$ corresponds to
all Pauli couplings between \emph{two} ground state excitons
$(0,0)$ and the two $(m,n)$ excitons; the second term of
$\mathcal{B}_{mn}^{(N)}$ corresponds to all Pauli couplings
between \emph{three} ground state excitons $(0,0,0)$ and the
three $(m,n,0)$ excitons; the third term couples
the $(0,0,0,0)$ excitons to the $(m,n,0,0)$ excitons, and so
on $\ldots$ 

\section{Density expansion of $<H>_N$}
Let us now return to the expectation value of the
hamiltonian $<H>_N$. Using Eqs (6,14,18,21), and the fact
that $\xi_{000m}^{\mathrm{dir}}$=0 (see Eq (7)), we find
that 
$<H>_N$ reads
\begin{equation}
<H>_N=N(E_0+\Delta)\ ,
\end{equation}
where the three first terms of $\Delta$ are given by 
\begin{eqnarray}
\Delta =
&-&(N-1)\,\frac{F_{N-2}}{F_N}\,\frac{S_1}{2}
+(N-1)(N-2)\,\frac{F_{N-3}}{F_N}\,S_2\nonumber
\\ &-&(N-1)(N-2)(N-3)\,\frac{F_{N-4}}{F_N}
\left(S_3+\frac{S'_3}{2}-\frac{S''_3}{2}\right)
+\cdots\ ,
\end{eqnarray}
the expressions of the various sums
$S_n$ in terms of
$\xi^{\mathrm{dir}}$'s and
$\lambda$'s being given below.                        

By using Eqs (4,10), it is easy
to see that the first sum $S_1$ reads in $\mathbf{r}$
space
\begin{eqnarray}
S_1&=&\sum_{mn}\xi_{00mn}^{\mathrm{dir}}\lambda_{mn00}
\nonumber
\\&=&\int
d\mathbf{r}_{e_1}\,d\mathbf{r}_{h_1}\,d\mathbf{r}_{e_2}\,
d\mathbf{r}_{h_2}\,\phi_0^\ast(e_1,h_1)\phi_0^\ast(e_2,h_2)
\nonumber
\\
&\times&(V_{e_1e_2}+V_{h_1h_2}-V_{e_1h_2}-V_{e_2h_1})\phi_0(e_1,h_2)\phi_0(e_2,h_1)\
,
\end{eqnarray}
which is nothing but $\xi_{0000}^{\mathrm{right}}\equiv \xi_{0000}^{\mathrm{exch}}$
defined in our previous works dealing with the matrix elements of $H$ in the
two-exciton subspace $^{(3)}$. This term is also the usual
exchange Coulomb term of the effective bosonic hamiltonian
for excitons $^{(2,8)}$. 

The second sum of Eq (23) involves three excitons. It reads
\pagebreak  
\begin{eqnarray}
S_2&=&\sum_{mnp}\xi_{00mn}^{\mathrm{dir}}\lambda_{mp00}\lambda_{n0p0}\nonumber
\\&=&\int
d\mathbf{r}_{e_1}\,d\mathbf{r}_{h_1}\,d\mathbf{r}_{e_2}\,d\mathbf{r}_{h_2}\,d\mathbf{r}_{e_3}\,d\mathbf{r}_{h_3}\,\phi_0^\ast(e_1,h_1)\phi_0^\ast(e_2,h_2)\phi_0^\ast(e_3,h_3)\nonumber
\\ & &\hspace{1cm}\times
[V_{e_1e_2}+V_{h_1h_2}-V_{e_1h_2}-V_{e_2h_1}]\,
\phi_0(e_1,h_2)\phi_0(e_2,h_3)\phi_0(e_3,h_1)\
.
\end{eqnarray}

The three sums of the third term of Eq (23) involves four
excitons. The first sum $S_3$ reads 
\begin{eqnarray}
S_3&=&\sum_{mnpq}\xi_{00mn}^{\mathrm{dir}}\lambda_{mp00}\lambda_{nqp0}\lambda_{00q0}\nonumber
\\&=&\int
d\mathbf{r}_{e_1}\,d\mathbf{r}_{h_1}\,d\mathbf{r}_{e_2}\,d\mathbf{r}_{h_2}\,d\mathbf{r}_{e_3}\,d\mathbf{r}_{h_3}\,d\mathbf{r}_{e_4}\,d\mathbf{r}_{h_4}\,\phi_0^\ast(e_1,h_1)\phi_0^\ast(e_2,h_2)\nonumber
\\& &\hspace{1cm}
\times\phi_0^\ast(e_3,h_3)\phi_0^\ast(e_4,h_4)[V_{e_1e_2}+V_{h_1h_2}-V_{e_1h_2}-V_{e_2h_1}]\nonumber
\\& &\hspace{1cm} 
\times\phi_0(e_1,h_2)\phi_0(e_2,h_3)\phi_0(e_3,h_4)\phi_0(e_4,h_1)\
.
\end{eqnarray}

The sum $S'_3$, defined as
$S'_3=\sum_{mnpq}\xi_{00mn}^{\mathrm{dir}}
\lambda_{mp00}\lambda_{nq00}\lambda_{00pq}$
, reads 
as Eq (26) except for the four $\phi_0$'s which
are replaced by
$$\phi_0(e_1,h_3)\phi_0(e_2,h_4)\phi_0(e_3,h_2)
\phi_0(e_4,h_1)\ .$$ 

The last sum $S''_3$, defined as
$S''_3=\sum_{mn}\xi_{00mn}^{\mathrm{dir}}\lambda_{m000}
\lambda_{n000}$, reads as Eq
(26) except for the four
$\phi_0$'s which are now replaced by
$$\phi_0(e_1,h_3)\phi_0(e_2,h_4)\phi_0(e_3,h_1)
\phi_0(e_4,h_2)\ .$$ This sum is in fact equal to zero as
easily seen by exchanging
$(\mathbf{r}_{e_2}\leftrightarrow\mathbf{r}_{h_2})$, and
$(\mathbf{r}_{e_4}\leftrightarrow\mathbf{r}_{h_4})$, in the
integral.

It is physically important to note that all these sums
contain the \emph{same Coulomb coupling between only two
excitons} made with $(e_1,h_1)$ and $(e_2,h_2)$. Couplings
between more than two excitons appearing in $S_{n³2}$,
result from many-body effects induced by the close-to-boson
character of the excitons.

The sums $S_1$, $S_2$, $S_3$, $S'_3$ are
calculated in the Appendix (see Eqs (40-43)). Although not
obvious at first, $S_1$, $S_2$ and $S_3+S'_3/2$ are in fact
real even if $\Phi_{\nu_0}$ is not, as necessary to have a
real energy change $\Delta$ induced by these Coulomb and
Pauli couplings.

$\Delta$ also contains density dependent terms through the
ratios
$F_{N-p}/F_N$ which differ from 1 due to the close-to-boson
character of the excitons. These ratios can be obtained from
the recursion relation (16), which also reads
\begin{equation}
\frac{F_N}{F_{N-1}}=1-(N-1)\,\sigma_2\,\frac{F_{N-2}}{F_{N-1}}
+(N-1)(N-2)\,\sigma_3\,\frac{F_{N-3}}{F_{N-1}}+\cdots
\end{equation}
As $\sigma_n$ is in $(a_x^3/\mathcal{V})^{n-1}$ (see Eq
(17)), we can iterate Eq (27) to obtain the ratio
$F_N/F_{N-1}$, for $N>>1$, as an expansion in powers of the
parameter
$\eta=Na_x^3/\mathcal{V}$. We get
\begin{equation}
\frac{F_N}{F_{N-1}}=1-N\sigma_2+N^2(\sigma_3-\sigma_2^2)+O(\eta^3)
\end{equation}
As for large $N$, $F_{N-p}/F_N\simeq(F_{N-1}/F_N)^p$, the
ratios appearing in Eq (23) are given by
\begin{equation}
\frac{F_{N-2}}{F_N}=1+2N\sigma_2+N^2(5\sigma_2^2-2\sigma_3)+O(\eta^3)
\ ,\hspace{0.2cm}\frac{F_{N-3}}{F_N}=1+3N\sigma_2+O(\eta^2)
\ ,\hspace{0.2cm}\frac{F_{N-4}}{F_N}=1+O(\eta)\ .
\end{equation}
By inserting Eq (29) into Eq (23), we
obtain the following expansion of the
energy change $\Delta$ in powers of
$\eta$ :
\begin{equation}
\Delta=-N\frac{S_1}{2}+N^2[S_2-\sigma_2S_1]+N^3\left[-S_3-\frac{S'_3}{2}+3\sigma_2S_2-\frac{S_1}{2}(5\sigma_2^2-2\sigma_3)\right]+O(\eta^4)
\end{equation}

Using Eqs (17,40-43) and putting
everything together, we finally find that
the hamiltonian expectation value in the
$N$-ground-state-exciton state expands in powers of
\linebreak
$\eta=Na_x^3/\mathcal{V}$ as
\begin{eqnarray}
<H>_N&=&NE_0\left(
1-\frac{13\pi}{3}\eta+\frac{73\pi^2}{20}\eta^2
-\frac{3517\pi^3}{210}\eta^3+O(\eta^4)\right)\nonumber
\\&=&NE_0\left(1-13.61\,\eta+36.02\,\eta^2-519.3\,\eta^3+O(\eta^4)
\right)
\end{eqnarray}
Let us again stress that terms in $\eta^n$ with $n\geq 2$
come from many-body effects induced by Pauli exclusion, since
Coulomb interaction enters $<H>_N$ at first order only.

\section{Conclusion}
We have calculated the expectation value $<H>_N$ of the
exact semiconductor hamiltonian in the
$N$-ground-state-exciton state $(B_0^\dag)^N|v>$, using the
commutation technique we recently introduced. Due to novel
many-body effects induced by the close-to-boson character of
the excitons, $<H>_N$ appears as an expansion in powers of
the density through $\eta=Na_x^3/\mathcal{V}$, Coulomb
interaction entering this quantity at first order only by
construction. Higher order terms in $\eta^n$, with $n\geq2$,
result from both, sophisticated exchange processes in which
the Coulomb interaction appears at first order only, and
purely Pauli many-body effects which make
$<v|(B_0)^N(B_0^\dag)^N|v>$ to differ from its exact boson
value $N$!.

From the result given in Eq (31), we see that the prefactors
of the expansion of $<H>_N$ are rather large so that
$\eta=Na_x^3/\mathcal{V}$ has to be much smaller than 1 for 
$<H>_N$ to be equal to the energy of $N$
non-interacting boson-excitons $NE_0$. In a
previous work $^{(7)}$, we have already shown
that, while the Mott criterion corresponds to
$\eta=Na_x^3/\mathcal{V}\simeq 1$ for the
electron-hole pairs to be bound in excitons, i.\
e.\ for the excitons to exist, the criterion for
the excitons to behave as bosons is more like
$100Na_x^3/\mathcal{V}\simeq 1$. As the $\eta$
expansion of $<H>_N$ is physically linked to
many-body effects induced by the close-to-boson
character of the excitons, it is after all
reasonable to find similar conditions for
excitons to behave as bosons and for the
hamiltonian expectation value to be equal to the one of
$N$ non-interacting boson-excitons.          
\newpage

\hbox to \hsize {\hfill APPENDIX
\hfill}
\vspace{1cm}

We start from the expressions of
$S_1$, $S_2$, $S_3$ and $S'_3$ in
$\mathbf{r}$ space given in
Eqs (24-26), and we rewrite them by using
the Fourier transforms of the exciton wave function and
Coulomb potential. For $\mathbf{Q}_0=\mathbf{0}$, we have
\begin{equation}
\phi_0(e,h)=\frac{1}{\mathcal{V}}\sum_{\mathbf{k}}
\Phi_{\nu_0}(\mathbf{k})\,e^{i\mathbf{k}.(\mathbf{r}_e
-\mathbf{r}_h)}\ ,
\end{equation}
\begin{equation}
\frac{e^2}{r}=\sum_{\mathbf{q}}
V_{\mathbf{q}}\,e^{i\mathbf{q}.\mathbf{r}}\ ,
\end{equation}
with $V_{\mathbf{q}}=4\pi e^2/\mathcal{V}q^2$ in 3D.

If we insert Eqs (32,33) into $S_1$ given in Eq (24), and we
perform the integrals over the $\mathbf{r}$'s, we get
\begin{equation}
S_1=2\sum_{\mathbf{k},\mathbf{k'}}V_{\mathbf{k}-\mathbf{k'}}
|\Phi_{\nu_0}(\mathbf{k})|^2\left[|\Phi_{\nu_0}(\mathbf{k'})|
^2-\Phi_{\nu_0}^\ast(\mathbf{k})\,\Phi_{\nu_0}
(\mathbf{k'})\right]\ .
\end{equation}
By using the Schršdinger equation for the exciton relative
motion,
\begin{equation}
(\hbar^2\mathbf{k}^2/2\mu -
\epsilon_{\nu_0})\Phi_{\nu_0}(\mathbf{k})-\sum_{\mathbf{k'}}
V_{\mathbf{k}-\mathbf{k'}}\,\Phi_{\nu_0}(\mathbf{k'})=0\
,
\end{equation}
with $\epsilon_{\nu_0}=E_0=-e^2/2a_x$, we can check that
$S_1$ is real.

If we now insert Eqs (32,33) into $S_2$ given in Eq (25),
and perform the integrals over the $\mathbf{r}$'s, we get
\begin{eqnarray}
S_2=\sum_{\mathbf{k},\mathbf{k'}}V_{\mathbf{k}-\mathbf{k'}}
\left[
2|\Phi_{\nu_0}(\mathbf{k})|^4\,|\Phi_{\nu_0}(\mathbf{k'})|
^2-|\Phi_{\nu_0}(\mathbf{k})|^4\,
\Phi_{\nu_0}^\ast(\mathbf{k})\,\Phi_{\nu_0}(\mathbf{k'})
\right.\nonumber
\\ \left. -\left(\Phi_{\nu_0}^{\ast
2}(\mathbf{k})\Phi_{\nu_0} (\mathbf{k})\right)
\left(\Phi_{\nu_0}^2(\mathbf{k'})\Phi_{\nu_0}^\ast
(\mathbf{k'})\right)\right]\ .
\end{eqnarray}
The second term of Eq (36) is real due to Eq (35), while the
last term is real by \linebreak
$(\mathbf{k}\leftrightarrow\mathbf{k'})$.

For the third order sums, we obtain in the same way
\begin{eqnarray}
S_3=\sum_{\mathbf{k},\mathbf{k'}}V_{\mathbf{k}-\mathbf{k'}}\,
|\Phi_{\nu_0}(\mathbf{k})|^4\left[\ 
2|\Phi_{\nu_0}(\mathbf{k})|^2\,|\Phi_{\nu_0}(\mathbf{k'})|^2
-|\Phi_{\nu_0}(\mathbf{k})|^2\,
\Phi_{\nu_0}^\ast(\mathbf{k})\,
\Phi_{\nu_0}(\mathbf{k'})\right.
\nonumber
\\ \left. -|\Phi_{\nu_0}(\mathbf{k'})|^2\,
\Phi_{\nu_0}(\mathbf{k})\,
\Phi_{\nu_0}^\ast(\mathbf{k'})\ \right]\ ,
\end{eqnarray}
\begin{equation}
S'_3=2\sum_{\mathbf{k},\mathbf{k'}}V_{\mathbf{k}-\mathbf{k'}}\,
|\Phi_{\nu_0}(\mathbf{k})|^4\left[\ 
|\Phi_{\nu_0}(\mathbf{k'})|^4
-|\Phi_{\nu_0}(\mathbf{k'})|^2\,
\Phi_{\nu_0}^\ast(\mathbf{k})\,\Phi_{\nu_0}
(\mathbf{k'})\ \right]\ .
\end{equation}
The energy shift $\Delta$ depending on $(S_3+S'_3/2)$ as
$S''_3=0$, we can check that this quantity is real due again
to Eq (35).

By inserting the 3D value of $\Phi_{\nu_0}(\mathbf{k})$,
namely
\begin{equation}
\Phi_{\nu_0}(\mathbf{k})=\frac{8\sqrt{\pi}}{(1+a_x^2k^2)^2}
\left(\frac{a_x^3}{\mathcal{V}}\right)^{1/2}\ ,
\end{equation}
into Eqs
(34,36,37,38), and by calculating the integrals
over $\mathbf{k}$ and $\mathbf{k'}$, we
obtain
\begin{equation}
S_1=\frac{26\pi}{3}\left(\frac{a_x^3}{\mathcal{V}}\right)
\,E_0\ ,
\end{equation}
\begin{equation}
S_2=\frac{2933\pi^2}{20}\left(\frac{a_x^3}{\mathcal{V}}
\right)^2\,E_0\ ,
\end{equation}
\begin{equation}
S_3=\frac{10795\pi^3}{2}\left(\frac{a_x^3}{\mathcal{V}}
\right)^3\,E_0\ ,
\end{equation}
\begin{equation}
S'_3=\frac{29601\pi^3}{28}\left(\frac{a_x^3}{\mathcal{V}}
\right)^3\,E_0\ ,
\end{equation}
with $E_0=-e^2/2a_x$.

\newpage

\hbox to \hsize {\hfill REFERENCES
\hfill}
\vspace{1cm}

(1) T. USUI, Prog. Theor. Phys. \underline{23}, 787 (1957).

(2) H. HAUG, S. SCHMITT-RINK, Prog. Quant. Elec.
\underline{9}, 3 (1984).

(3) M. COMBESCOT, O. BETBEDER-MATIBET, Cond-mat/0106346
(18 June 2001), to appear in Europhysics Letters.

(4) M. COMBESCOT, O. BETBEDER-MATIBET, Cond-mat/0201554 (30
Jan 2002), submitted to Europhysics Letters.

(5) For a review, see A. KLEIN, E. R. MARSHALEK, Rev. Mod.
Phys. \underline{63}, 375 (1991).

(6) O. BETBEDER-MATIBET, M. COMBESCOT, submitted to Eur.
Phys. J. B.

(7) M. COMBESCOT, C. TANGUY, Europhysics Letters
\underline{55}, 390 (2001).

(8) L. V. KELDYSH, A. N. KOZLOV, Sov. Phys. JETP
\underline{27}, 521 (1968).

\newpage

\hbox to \hsize {\hfill FIGURE CAPTIONS
\hfill}
\vspace{0.5cm}

Fig. (1)

(a) Direct Coulomb interaction $\xi_{mnij}^{\mathrm{dir}}$
between two excitons which are scattered from the $(i,j)$
states to the $(m,n)$ states, the "in" and "out" excitons
being made with the same electron-hole pairs.

(b) Coupling between $(i,j)$ and $(m,n)$ excitons induced by
the fact that two electrons and two holes can be coupled in
two different ways to form two excitons.

\vspace{0.5 cm}

Fig. (2)

Overlap $\mathcal{A}_m^{(N)}$, as defined in Eq (12),
between two $N$-exciton states, when one of the $N$ excitons
is in a state $m$, instead of the ground state $0$.

(a) Diagrammatic representation of the integral equation
(13) which links $\mathcal{A}_m^{(N)}$ to
$\left(\mathcal{A}_p^{(N-1)}\right)^\ast$. The cross
represents the $\lambda_{mp00}$ exchange process.

(b) Density expansion of $\mathcal{A}_m^{(N)}$ as given in
Eq (15). The quantity $F_N$, defined in Eq (14), differs
from 1 because the excitons are not exact bosons.

\vspace{0.5 cm}

Fig. (3)

Overlap $\mathcal{B}_{mn}^{(N)}$, as defined in Eq (18),
between two $N$-exciton states when two of the $N$ excitons
are in the states $(m,n)$ instead of the ground state 0.

(a) Diagrammatic representation of the integral equation
(20) which links $\mathcal{B}_{mn}^{(N)}$ to 
$\left(\mathcal{A}_q^{(N-2)}\right)^\ast$ and
$\left(\mathcal{B}_{pq}^{(N-2)}\right)^\ast$. The cross of
the first line corresponds to $\lambda_{mn00}$ while the two
crosses of the last line correspond to $\lambda_{nq00}$ and
$\lambda_{mp00}$.

(b) Density expansion of $\mathcal{B}_{mn}^{(N)}$ as given
in Eq (21). $F_N$ is defined in Eq (14).

\vspace{0.5 cm}

Fig. (4)

(a) The sum $S_1$ defined in Eq (24) with one exchange
process
$\lambda_{mn00}$ which transforms the $(0,0)$ excitons into
$(m,n)$ excitons, followed by one direct Coulomb interaction
$\xi_{00mn}^{\mathrm{dir}}$ which scatters these $(m,n)$
excitons back into the ground state $(0,0)$.

(b) The sum $S_2$, defined in Eq (25), with two exchanges and
one direct Coulomb scattering. 

(c) (d) (e) The sums $S_3$,
$S'_3$ and $S''_3$, defined in Eq (26) and below, with three
exchanges and one direct Coulomb scattering.

\end{document}